\documentclass[referee]{raa_rb}
\usepackage{graphicx,times,diagbox}
\usepackage{natbib}
\usepackage{amssymb,amsmath}
\usepackage{threeparttable}

\begin{document}
   \title{Photometric Analysis of the eclipsing Polar MN Hya}
   \volnopage{Vol.0 (20xx) No.0, 000--000}
   \setcounter{page}{1}
   
   \author{Qi-Shan Wang\inst{1,2,3}\thanks{email:wangqs@ynao.ac.cn},
           Sheng-Bang Qian\inst{1,2,3}, Zhong-Tao Han\inst{1,2,3}, Miloslav Zejda\inst{4}, Eduardo Fern\'andez-Lajus\inst{5,6}, 
           \and Li-Ying Zhu\inst{1,2,3} 
   }
   
   \institute{Yunnan Observatories, Chinese Academy of Sciences (CAS), P. O. Box 110, 650216 Kunming, China.\\
   \and
   Key Laboratory of the Structure and Evolution of Celestial Objects, Chinese Academy of Sciences, P. O. Box 110, 650216 Kunming, China.\\
   \and
   University of Chinese Academy of Sciences, Yuquan Road 19\#, Sijingshang Block, 100049 Beijing, China.\\
   \and
   Department of Theoretical Physics and Astrophysics, Masaryk University, Kotl\'ar\u sk\'a 2, Brno 611 37, Czech Republic.\\
   \and
   Facultad de Ciencias Astron\'omicas y Geof\'isicas, Universidad Nacional de La Plata, Paseo del Bosque s/n, 1900, La Plata, Pcia. Bs. As., Argentina.\\
   \and
   Instituto de Astrof\'isica de La Plata (CCT La plata - CONICET/UNLP), Argentina.\\
\vs \no 
   {\small Received 2018 month day; accepted 2018 month day}
}

\abstract{As an eclipsing polar with 3.39 hrs orbital period, MN Hya was going through state change when we observed it during 2009-2016. 10 new mid-eclipse times, along with others obtained from literature, allow us to give a new ephemeris. The residuals of linear fit show that period decreased during the phase of state change. It means angular momentum was lost during this phase. The X-ray observation indicates the mass accretion rate as about $3.6\times10^{-9}M_{\odot}yr^{-1}$. The period decrease gives that at least  60 percent of mass being transfered from secondary was lost, maybe in form of the spherically symmetric stellar wind. In high state, the data shows the intensity of the flickering reduced when system had higher accretion rate, and that flickering sticks out with primary timescale about 2 minutes, which implies the position of the threading point as about 30 radius of the white dwarf above the surface of it. The trend of light curves of the system in high state follows that of low state for a large fraction of phase interval from phase 0 to phase 0.4 since which the cyclotron feature is visible, and the primary intensity hump of light curves near phase 0.7 when the system is in high state did not appear on the curve when it is in low state. Those facts contradict the predictions of the two-pole model.
}
   \authorrunning{Qi-Shan Wang et al.}
   \titlerunning{Photometric Analysis of MN Hya}
   \maketitle

\section{Introduction}\label{sect:intro}
MN Hya was detected as the optical counterpart of the X-ray source RX J0929.1-2404 and identified as a polar due to the spectral characteristics \citep{Sekiguchi1994RXJ}. As one of the subtypes of cataclysmic variables (CVs), polars are interacting binaries in which mass transfers from a dwarf secondary star to the primary, a magnetic ($\sim10-200MG$) white dwarf (WD), through Roche lobe overflow. The magnetic field is sufficiently high to prevent the formation of an accretion disc and channels the accretion stream at threading point, then guides the flow plunging into the white dwarf directly around the magnetic pole \citep{Hellier2001Cataclysmic}. If only one accretion region is visible, we call the behavior as one-pole behavior. In contrast to two-pole behavior, two accretion regions appear alternatively. Because of the variation of mass transfer rate, polars switch irregularly between high state and low state. In high state, polars present fast light intensity variation which is known as flickering. Later, several observations of MN Hya, including photometry, spectrometry, X-ray, and polarimetry, let some authors classify it as a two-pole system. The primary accretion pole, below the orbital plane, appears between phase $\sim0.5-1$ and causes the intensity peak near phase 0.7 while the secondary accretion pole contributes its emission all the way \citep{Buckley1998Polarimetry}. Also in X-rays, MN Hya shows a prominent pre-eclipse dip around phase 0.9, but it is not detected in the optical passband with the same depth and phase interval \citep{Buckley1998Xray}. Moreover, MN Hya is an eclipsing system with a high orbital inclination of $\sim75^{\circ}$ \citep{Buckley1998Polarimetry}. The eclipsing nature provides a good opportunity to ascertain its geometric structure and evolutionary state, and also to search for circumbinary planets. Since 2009 our group started to detect the secular evolution of CVs and the extrasolar planets around them by the eclipse timing method (e.g. \citealt{Qian2009Orbital,Qian2010DETECTION,Qian2011Detection,Qian2015Long,Qian2016Rapid}; \citealt{Dai2009Evidence,Dai2010Orbital}; \citealt{Han2015An,Han2016Physical,Han2017Cyclic,Han2017Double,Han2017WZ}). 

   In this paper we analysis the photometric observations of MN Hya. We give a revised ephemeris and find that the orbital period decreased during the state change. The light curve (LC) of low state presents one hump near phase 0.2 and those of high state show obvious flickering. The paper is organized in the following way. In Sect.~\ref{sect:observ} the observation of the datasets is described. And a detail analysis is given in Sect.~\ref{sect:analysis}. Our main results are summarized in Sect.~\ref{sect:summary}.

\section{Observations}\label{sect:observ}
\begin{figure}
   \centering
   \includegraphics[width=14.0cm, angle=0]{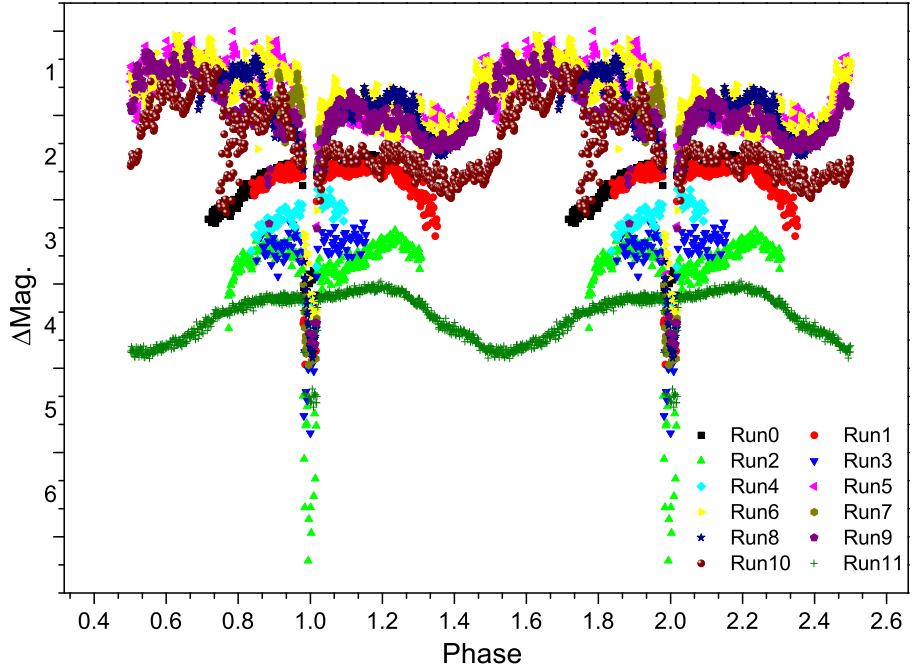}
   \caption{All light curves detected during 2009-2016. The ordinate is the differential magnitude of the target star with respect to the comparison star. The labels indicate the observation IDs used in this paper. More information is given in Table \ref{tab1}.}
   \label{Fig.V_K}
   \end{figure}

MN Hya was observed since 27, November 2009 by the Roper Scientific, Versarray 1300B camera with a thinned EEV CCD36-40 de 1340 $\times$ 1300 pixel CCD chip, attached to the 2.15-m "Jorge Sahade" telescope (JST) at Complejo Astron\'omico E1 Leoncito (CASLEO), San Juan, Argentina. Later, this binary was monitored during 2012-2016 by using the Danish 1.54-m Telescope at ESO La Silla, Chile. MN Hya was detected photometrically for 12 times with no filter used mostly, apart from one observation through V filter in 17, January 2016. The CCD images were reduced with the help of the aperture photometry package of IRAF \citep{Tody1993IRAF}. Nearby non-variable stars were chosen as comparison star and check star (the stars marked C and B respectively in figure 2 of \cite{Sekiguchi1994RXJ}), so two sets of different magnitudes of each observation, i.e. the magnitudes of the object and the companion star with respect to that of companion star and check star respectively, were obtained and used for the analysis of the characteristics of the light curve of the object.
An overview of the observation of MN Hya is given in Table~\ref{tab1}. The table lists the information about an observation ID used in this paper, the data of the observation, the state of the object, the time resolution, the telescope and the filter used, and the total phase coverage.   
Fig. \ref{Fig.V_K} displays all LCs obtained in 2009-2016. It is lucky of us to detect the process of the state change which gives us enough information to know the characteristics of this phase.

\begin{table}[!ht]
\bc
\begin{minipage}[]{100mm}
\caption[]{Log of photometric observation of MN Hya. The meaning of the entries are given in context.\label{tab1}}\end{minipage}
\setlength{\tabcolsep}{5pt}  
\small
\begin{threeparttable}
\begin{tabular}{ccccccc}
  \hline\noalign{\smallskip}
Obs-ID & Date & accretion state & resolution(s) & Telescopes & Filter & Phase Coverage\tnote{a}\\
  \hline\noalign{\smallskip}
Run0&20091127&intermediate&23&2.15m&N&0.72-1.31\\
Run1&20091128&intermediate&23&2.15m&N&0.84-1.35\\
Run2&20100115&intermediate&28&2.15m&N&0.77-1.31\\
Run3&20101227&intermediate&40&2.15m&N&0.85-1.15\\
Run4&20101230&intermediate&40&2.15m&N&0.84-1.09\\
Run5&20121225&high        &16&1.54m&N&0.82-1.82\\
Run6&20130406&high        &16&1.54m&N&0.18-1.43\\
Run7&20131223&high        &14&1.54m&N&0.89-1.08\\
Run8&20150129&high        &16&1.54m&N&0.69-1.43\\
Run9&20160101&high        &21&1.54m&N&0.03-1.08\\
Run10&20160117&high       &22&1.54m&V&0.17-1.38\\
Run11&20160310&low        &17&1.54m&N&0.89-1.95\\
  \noalign{\smallskip}\hline
\end{tabular}
      \begin{tablenotes}
        \footnotesize
        \item[a] Phase is calculated according to the ephemeris of equation (\ref{Eq.1}).
      \end{tablenotes}
\end{threeparttable}
\ec
\end{table}

\section{Analysis} \label{sect:analysis}
 \subsection{Period Change} \label{subsect:period}
\begin{figure}
   \centering
   \includegraphics[width=14.0cm, angle=0]{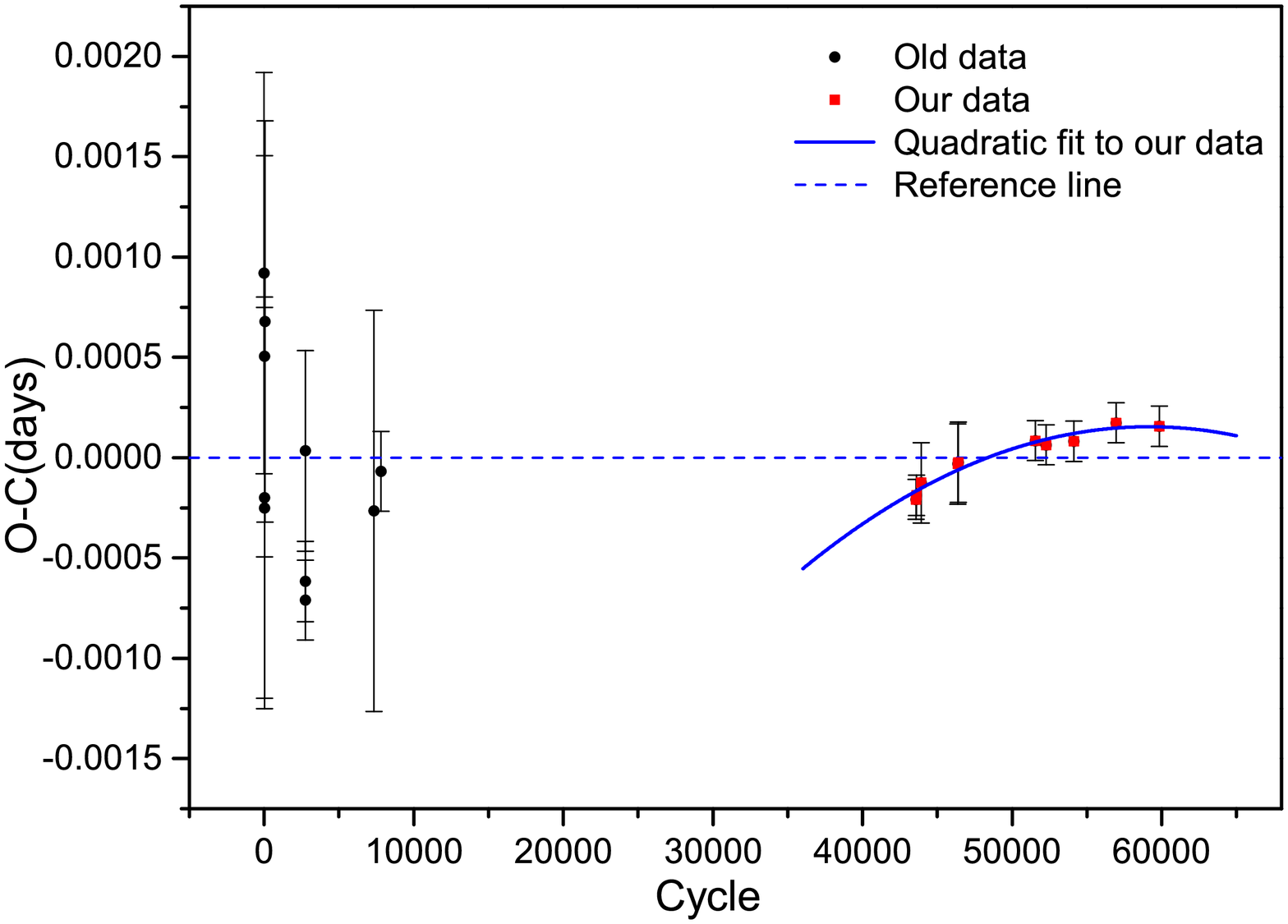}
   \caption{The residuals of the linear fit to the mid-eclipse times with the reference line for convenience. The new data overlaid by fitted parabola  shows period decrease.}
   \label{Fig.O-C}
   \end{figure}

    From our photometric observation, 10 mid-eclipse times, the average of eclipse ingress and egress, were obtained, and a linear least-squares fit to the timings gives the following eclipsing ephemeris:
    \begin{equation}
    \label{Eq.1}
    HJD_{min}=2457458.62831(4) + 0.1412437990(31)E
    \end{equation}
    Fig.~\ref{Fig.O-C} shows the residual (O-C, observed minus calculated) times of the linear fit. The O-C curve indicates that the orbital period decreased during the state change. Because the mid-eclipse time is that of the occultation of the primary by the secondary \citep{Schmidt2001HSTSpectroscopy} which can not be effected by the shift of the accretion spot due to the variation of accretion rate, the O-C curve of the system will give the information of orbital evolution. The new data has higher precision, it makes the possibility of analysis the orbital variation. The period decreased as $-2.4\times10^{-12}dcyc^{-1}$ with the error of $0.7\times10^{-12}dcyc^{-1}$. In the angular momentum conservation case the orbital period increases when mass transfers from the less massive secondary to the more massive primary WD star, which makes the O-C curve as a parabola with a positive quadratic term. So we infer that the mass transfers from the secondary to the primary accompanied by intense mass loss which carries angular momentum from the system and decreases the orbital period. 

    In oder to estimate the mass loss fraction from the system during the high state, we need to know the accretion luminosity. The period ($P\approx3.39h$) implies that the secondary has mass $M_{2}=0.22M_{\odot}$  and radius  $R_{2}=0.32R_{\odot}$ (\citealt{Knigge2010Erratum}).  Considering of the inclination of system $i=75^{\circ}$ and the phase width of eclipse about 0.036,  we deduce the mass ratio ($q=M_{2}/M_{1}$) as about 0.38 \citep{Horne1985Images} which gives the primary with mass $M_{1}=0.58M_{\odot}$ and radius $R_{1}=0.88\times10^{9}cm$ \citep{Nauenberg1972Analytic}. The $0.1-2.4$k$eV$ luminosity is about $9\times10^{32} erg s^{-1}$ \citep{Ramsay1998Spectroscopic}, with the addition of the soft X-ray excess ($\frac{L_{bb}}{L_{brems}+L_{cycs}}$) $\sim$ 10 \citep{Buckley1998Xray} and assuming $L_{brems}=L_{cycs}$, we derive that the luminosity of accretion is about $L_{acc}=2\times10^{34} erg s^{-1}$ which gives the mass accretion rate of $\dot{M_{1}}=3.6\times10^{-9}M_{\odot}yr^{-1}$. 
    Combine
   \begin{equation}\label{Eq.2}
    \frac{\dot{J}}{J}=\frac{\dot{M}_{1}}{M_{1}}+\frac{\dot{M}_{2}}{M_{2}}-\frac{1}{2}\frac{\dot{M}}{M}+\frac{1}{2}\frac{\dot{a}}{a},
   \end{equation}
with Kepler's law and assuming mass loss rate $\dot{M}=\alpha\dot{M_{2}}$, we can write
   \begin{equation}\label{Eq.3}
    \frac{\dot{J}}{J}=\left(1+\frac{1}{q(\alpha-1)}-\frac{1}{3}\frac{\alpha}{(1+q)(\alpha-1)}\right)\frac{\dot{M}_{1}}{M_{1}}+\frac{1}{3}\frac{\dot{P}}{P},
   \end{equation}
    where $M=M_{1}+M_{2}$ is the mass of the system and $J=M_{1}M_{2}~\sqrt[]{Ga/M}$ is the angular momentum of system. 
    $\dot{J}/J$ term consists of the angular momentum loss due to mass lost from system and redistribution of mass between two components. The first factor (denoted as loss-term) can be expressed by
  \begin{equation}\label{Eq.4}
    \left(\frac{\dot{J}}{J}\right)_{loss}=\frac{1+q}{q}\frac{\alpha}{\alpha-1}\frac{\dot{M_{1}}}{M_{1}}\left(\frac{r_{de}}{a}\right)^{2},
  \end{equation}
in terms of the effective decoupling position $r_{de}$: 
   \begin{equation}\label{Eq.5}
     \left(\frac{r_{de}}{a}\right)^{2}=\left(\frac{r_{1}}{a}\right)^{2}+\left(\frac{r_{2}}{a}\right)^{2}+2\frac{r_{1}}{a}\frac{r_{2}}{a}cos\Delta\theta,
   \end{equation}
where $r_{1}$ is the line length from one of components to the mass center of system (COM), $r_{2}$ is that from the decoupling point to this component, and $\Delta\theta$ is the angel between the two lines assuming 0 degree corresponding to the decoupling point in the opposite direction to COM. And the second (denoted as syn- term) can be calculated using
  \begin{equation}\label{Eq.6}
   \left(\frac{\dot{J}}{J}\right)_{syn}=\frac{2\pi\dot{M_{1}}a^{2}}{JP}\left[\left(\frac{l_{1}}{a}\right)^{2}-\left(\frac{q}{1+q}\right)^{2}\right]=\frac{1+q}{q}\frac{\dot{M_{1}}}{M_{1}}\left[\left(\frac{l_{1}}{a}\right)^{2}-\left(\frac{q}{1+q}\right)^{2}\right],
  \end{equation}
  with 
    \begin{equation}\label{Eq.7}
      l_{1}=a(0.5-0.227logq)-\frac{qa}{1+q},
    \end{equation}
giving the distance of the Lagrangian radius ($R_{L1}$, \citealt{Plavec1964Tables}) to COM.

    Substituting the parameters into the equations above deduces
    \begin{equation}\label{Eq.8}
      \left(\frac{r_{de}}{a}\right)^{2}\sim\frac{1.12}{\alpha}-0.46.
    \end{equation}
    The constraint $\alpha<1$ gives $r_{de}/a>0.81$ which implies the mass lost from the secondary. Note that the shortest length scale of the decoupling point to the COM only depends on the mass ratio, i.e., it is independent of the variation of period, mass accretion rate or other physical parameters. The maximal length scale of system about $1.2a$, corresponding to the distance of Lagrangian point L2 to COM, shows the minimal mass loss fraction $\alpha$ as about $60\%$, which reveals that the mass transfer rate from the secondary is about $6\times10^{-9}M_{\odot}yr^{-1}$ at least. 
    The Alf\'ven radius, considering the equivalence of ram pressure of flow and the magnetic pressure of field, is given by \citep{Frank2002Accretion} 
      \begin{equation}\label{Eq.9}
          r_{A}= 2.9 \times 10^{8} M_{1}^{-1/7} R_{6}^{10/7} L_{37}^{-2/7} B_{12}^{4/7},
       \end{equation}
from which we can estimate Alf\'ven radius $r_{A}$ is about $1.02a$. The magnetosphere centered the primary can cover about 42\% surface area of the secondary, which can cause the mass loss fraction of about $58\%$ if the secondary ejects matter symmetrically, such as the global intense wind. This kind of activity can produces mass loss fraction of about 58\% that is very close to the mass loss fraction $\alpha\sim60\%$ if mass is lost through the $L_{2}$. Whatever, we can see that the mass loss fraction is very large and any model not taking mass loss into account should be unreality.
 
\begin{table}[!ht]
\bc
  \begin{minipage}[]{100mm}
      \caption[]{All mid-eclipse times of MN Hya 
      \label{tab2}}
  \end{minipage}
\setlength{\tabcolsep}{5pt}
\small
\begin{threeparttable}
 \begin{tabular}{cccccc}
  \hline\noalign{\smallskip}
Min.(HJD) & Cycle & Error (days) & Calculated & O-C (days) & Ref.\tnote{*}\\
  \hline\noalign{\smallskip}
2449007.58900&0    &0.001  &2449007.58808&0.00092 &(1)\\
2449009.42400&13   &0.001  &2449009.42425&-0.00025&(1)\\
2449009.56600&14   &0.001  &2449009.56549&0.00051 &(1)\\
2449010.55400&21   &0.001  &2449010.55420&-0.00020&(1)\\
2449013.52100&42   &0.001  &2449013.52032&0.00068 &(1)\\
2449397.42100&2760 &0.0005 &2449397.42097&0.00003 &(1)\\
2449397.56150&2761 &0.0002 &2449397.56221&-0.00071&(1)\\
2449398.55030&2768 &0.0002 &2449398.55092&-0.00062&(1)\\
2450045.02352&7345 &0.0010 &2450045.02378&-0.00026&(1)\\
2450109.85462&7804 &0.0002 &2450109.85469&-0.00007&(1)\\
2455163.41637&43583&0.0001 &2455163.41657&-0.00021&(2)\\
2455164.40509&43590&0.0001 &2455164.40528&-0.00019&(2)\\
2455212.28680&43929&0.0002 &2455212.28693&-0.00013&(2)\\
2455558.33420&46379&0.0002 &2455558.33424&-0.00003&(2)\\
2455561.30033&46400&0.0002 &2455561.30036&-0.00002&(2)\\
2456287.71730&51543&0.0001 &2456287.71721&0.00008 &(2)\\
2456389.69530&52265&0.0001 &2456389.69524&0.00006 &(2)\\
2456650.85510&54114&0.0001 &2456650.85502&0.00008 &(2)\\
2457052.83505&56960&0.0001 &2457052.83487&0.00017 &(2)\\
2457458.62847&59833&0.0001 &2457458.62831&0.00016 &(2)\\
  \noalign{\smallskip}\hline
\end{tabular}
      \begin{tablenotes}
        \footnotesize
        \item[*] References: (1) \cite{Sekiguchi1994RXJ}; (2) Our observations.
      \end{tablenotes}
\end{threeparttable}
\ec
\end{table}

 \subsection{The characteristic of light curve}\label{subsect:characteristic}
 Now we describe the characteristics of the LCs more carefully and analysis the reasons for corresponding variation. And the analysis of the rapid light intensity is given in Subsection~\ref{subsect:flickering}.\\
 
\begin{figure}
   \centering
  \includegraphics[width=14.0cm, angle=0]{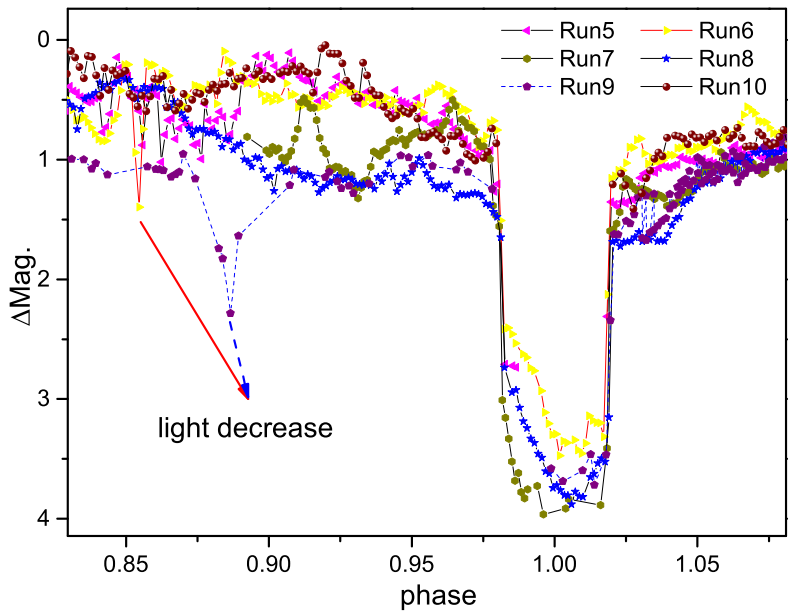}
   \caption{A detailed drawing of high-state LCs. The arrows point the position of per-eclipse light decrease.}
   \label{Fig.high_state}
   \end{figure}

    (a) Fig.~\ref{Fig.high_state} shows the details of high-state LCs. Some LCs show the conspicuous pre-eclipse light decrease, such as near phase 0.85 in Run6 and phase 0.89 in Run9. It is reminiscence of the X-ray pre-eclipse dip but with less phase extension. Also the eclipse profile variates significantly, especially that of Run8. The intensity during the eclipse of Run8, after a possibly flat base ahead of the slow decrease, rises slowly up, which indicates that some part of the accretion flow became brighter and system had higher mass transfer rate. The smoothed increase and decrease of light during eclipse mean that the fast light intensity variation is not due to the accretion flow. All of these changes presented in LCs reflect a significant variation of the mass accretion rate during high state.\\
    
\begin{figure}
   \centering
  \includegraphics[width=14.0cm, angle=0]{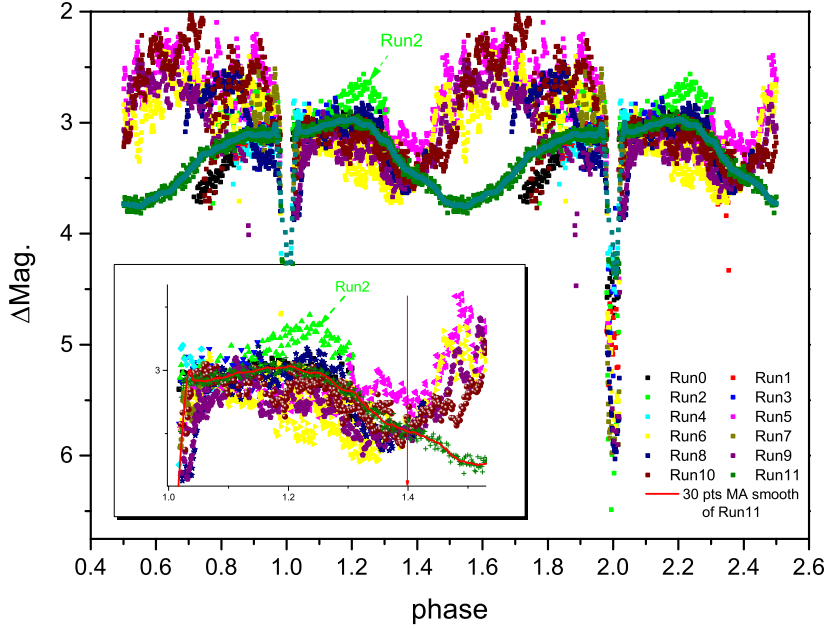}
   \caption{All LCs shifted by the average magnitude between 0.07-0.09. Inserted graph shows the the outline of LCs during phase 0-0.5 with the vertical line indicating the phase at which the deviation starts.} 
   \label{Fig.V-K_shift}
   \end{figure}

    (b) In Fig.~\ref{Fig.V-K_shift}, all data were shifted by the average magnitude between phase 0.07-0.09. Refer to Fig.~\ref{Fig.V_K}, we can find most high-state LCs follow the LC of Run11 very well during phase $\sim0-\sim0.4$, except those of Run5 and Run6. The reason of their deviation, we think, is due to instability of the system at the beginning of high state, such as Run2 which gives the light intensity variation of system at start of leaving low state. The large phase interval $\phi_{f}\sim0.4$ suggests that the only reasonable source giving rise to the hump near phase 0.2 is the WD itself. So from the shifting value of LCs, we can deduce the optical temperature relation of WD between low and high state. Assuming black body radiation and using the magnitude formula, we can get
         \begin{equation}\label{Eq.10}
            m_{H}-m_{L}=-2.5log\frac{f_{H}}{f_{L}}=-2.5log(\frac{S_{H}\bar{T}_{H}^{4}}{S_{L}\bar{T}_{L}^{4}}),
         \end{equation} 
where $m,S,\bar{T}$ is respectively the magnitude, projected surface area and mean temperature of the system, and L, H denote low and high state respectively. Considering, during phase 0-0.4, the dominating optical radiation of the WD and the spherical symmetry of that, the projected surface area is constant at certain phase, then we find
         \begin{equation}\label{Eq.11}
            m_{H}=m_{L}-10log(\frac{\bar{T}_{H}}{\bar{T}_{L}})=m_{L}+\Delta Mag,
         \end{equation}
where $\Delta Mag$ is the amount of translation for high-state LCs to overlap that of Run11. With $\Delta Mag=-2.21$, we find $\bar{T}_{2}\approx1.66\bar{T}_{1}$. If the temperature of the WD in low state is about 9000K \citep{Ramsay2004XMM}, we deduce the optical temperature of the WD in high state as about 15000K. Also as shown in the inserted graph of Fig.~\ref{Fig.V-K_shift}, the LCs of high state wander off that of Run11 near phase 0.4 since which the cyclotron features visible \citep{Ramsay1998Spectroscopic} and the accretion region shows up. Both the large $\phi_{f}$ during phase 0-0.4 and the visible cyclotron features during phase 0.4-1 suggest that it is an one-pole system.\\

   
\subsection{Flickering}\label{subsect:flickering}
  Even the LCs of intermediate state has less phase coverage than others, we can find the flickering only sticks out in high state, which indicates that the flickering accompanies high accretion rate and reminds us that transfer of material from the secondary is turbulent. In order to make clear the characteristics of the optical flickering of MN Hya, we analysis the spectrum of the LCs with enough phase coverage, the timescale of light intensity variation and the magnitude of the flickering in different parts of these LCs.
  
\begin{figure}
   \centering
   \includegraphics[width=14.0cm, angle=0]{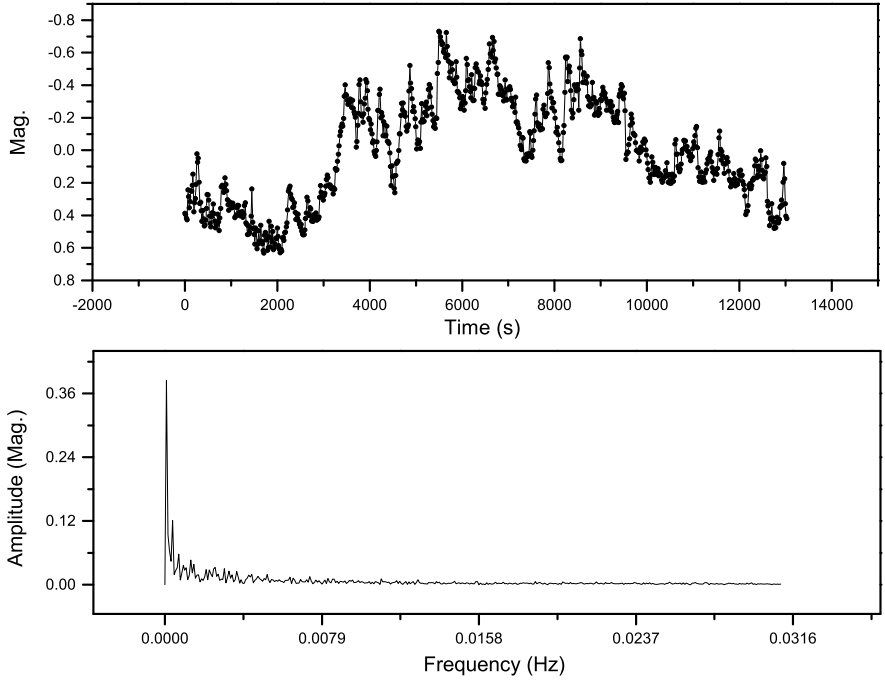}
   \caption{The upper panel shows the time series of Run6 and the lower panel gives its spectrogram determined by discrete Fourier transform. The dominate peak corresponds to the time interval of data.}
   \label{Fig.run6_dft}
   \end{figure}

  The data of each run were subjected to period searches using a discrete Fourier transform (DFT) and the phase dispersion minimization (PDM) methods (\citealt{Stellingwerf1978Period}, or \citealt{Dai2016Cataclysmic,Dai2017Quiescent} as reference). Fig.~\ref{Fig.run6_dft} shows the spectrogram of the time series of Run6 determined by DFT. The dominate peak corresponds to the time interval of the data and others with amplitude greater than 0.05 are the harmonics of this time. PDM method provides the similar result. This result manifests the rapid time variation is random, which is same as the characteristic of X-ray \citep{Buckley1998Xray}. But the system shows obvious evidence for brightness variations on timescale of minutes, the explains maybe that there are some oscillations which last for only few phase interval and then die away to be replaced by other oscillations with a different period or at different phase.
  
\begin{figure}
   \centering
  \includegraphics[width=14.0cm, angle=0]{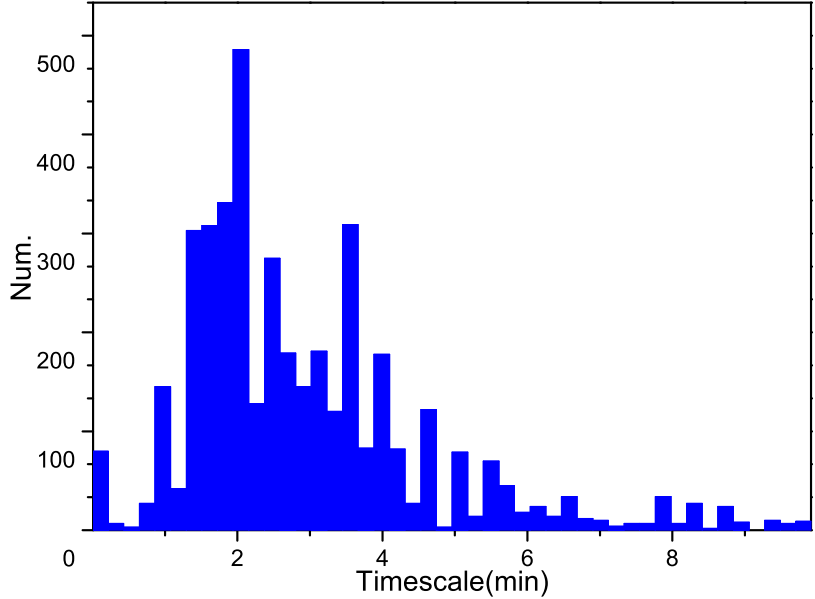}
   \caption{Histogram of the variation timescales of all data. The dominating timescale is about 2 min.} 
   \label{Fig.all_bin}
\end{figure}

   In order to find the characteristic timescale, we need to count up the number of different timescales. The method employed to deal with the data is described below. First,  in order to reduce the effect of variation of atmospheric seeing which can cause saw-toothed variation in adjacent sequence, we use moving average with 3 points to smooth the raw data of each run; then mark all the local extrema (maxima and minima), using which we give the histogram of the timescales between the extrema of the same type. As shown in Fig.~\ref{Fig.all_bin}, we find the dominating variation timescale as about 2 minutes, which is typical timescale in X-ray of other systems, such as EF Eri \citep{Patterson1981A} (period 6 min), VV Pup \citep{Maraschi1984Preliminary} (period 3 min), ST LMi \citep{Beuermann1985X} (period 1 min). If these variations occur at the threading point where the matter is channeled by magnetic filed, the timescale for Alfv\'en wave caused by instable accretion to cross the magnetosphere is
    \begin{equation}
    \label{Eq.12}
    P_{ins}(r) = 2 \times 10^{-3} r_{8}^{11/4} L_{34}^{1/2} f_{-2}^{-1/2} M_{1}^{-3/4} R_{8}^{-3/4} B_{7}^{-1} s,
    \end{equation}
where $r_{8}$ is the radius (in units of $10^{8} cm$, denoted as $r_{t}$ below) of the threading point, $L_{34}$ is the luminosity in units of $10^{34} erg s^{-1}$, $f$ is the fraction of the stellar surface where the accretion matter flows, $M_{1}$ is the WD mass in units of solar mass, $R_{8}$ is the radius of the WD in units of $10^{8} cm$, and $B_{7}$ is the surface magnetic field strength in units of $10^{12}G$. Adopting $f\approx 0.004$ \citep{Schmidt2001HSTSpectroscopy}, $B=40MG$ \citep{Buckley1998Polarimetry}, and other parameters mentioned earlier, we find the $r_{t}$ estimated for the timescale of 2 min as $2.8\times10^{10}cm$ or $30R_1$ ($0.62R_{L1}$). Comparing with the Alfv\'en radius, this length scale is much smaller, but is similar with that of AR UMa ($r_{t}\approx0.6-0.7R_{L1}$) shown on Doppler tomograms \citep{Hoard1999Accretion}, although the latter system has highest magnetic field. If the optical flickering is caused by the wobbling of threading point, the impact of the magnetic field strength on the position of that is very little.
    
    In order to quantify the flickering intensity, we remove the trend from the raw data using 30 points moving average method and calculate the standard deviation of the detrending data as the flickering intensity.
    Table~\ref{tab3} shows the flickering intensity corresponding to different phase interval, considering the two-pole model where Source1, Source2 and Total represent phase 0.03-0.4, 0.4-0.97 and all phase coverage, respectively. From this table, we can find the flickering intensity of Source1 is less than that of Source2 and that the flickering intensity reduces when the high-state system has higher accretion rate, such as Run8.

\begin{table}[!ht]
\bc
\begin{minipage}[]{100mm}
\caption[]{The flickering intensity of different parts given in table is the standard deviation of the detrending data of all phase, phase $\sim0.03-0.4$, and phase$\sim0.4-0.97$, those parts are denoted as Total, Source1, and Source2, respectively.\label{tab3}}\end{minipage}
\setlength{\tabcolsep}{5pt}  
\small
\begin{threeparttable}
\begin{tabular}{ccccccc}
  \hline\noalign{\smallskip}
\diagbox{Source}{Intensity (Mag.)}{Data}
       &Run5 &Run6 &Run8 &Run9 &Run10&Run11\\
\hline\noalign{\smallskip}
Total  &0.117&0.104&0.065&0.085&0.142&0.033\\
Source1&0.072&0.082&0.058&0.063&0.063&0.029\\
Source2&0.137&0.121&0.072&0.100&0.181&0.035\\
  \noalign{\smallskip}\hline
\end{tabular}
\end{threeparttable}
\ec
\end{table}

\section{Discussion and Summary}\label{sect:summary}
 From the discovery mentioned above, we can describe the mass transfer process in MN Hya. The magnetosphere covers  42\% surface area of the secondary. Convection of the secondary can entangle the magnetic field which will be wound up, increasing their tension and stored energy until it reconnects to dissipate energy and cast matter globally, maybe in forms of intense stellar wind.  This kind of casted matter with high velocity will pass through the Alf\'ven radius freely. When matter flow reaches the threading point which has hight of $\sim30R_{1}$ above the surface of WD, it will be threaded by the magnetic field. As the system is in high state, the accretion rate still variates which can trigger the instability of the accretion column, and then a Alf\'ven wave will arise from the column and travel through accretion stream to the threading point. This wave causes the instability of the threading point whose wobbling could modulate the mass accretion rate then give rise to fluctuation which causes the flickering. \\

   In summary, the optical observations of the eclipsing polar MN Hya for 12 times during 2010-2016, with the addition of other observations, enable us get the following results:\\
1. A revised ephemeris is given based on 10 more mid-eclipse times, along with others obtained from literature and the quadratical fit to new data shows the period decreased during the state change as $-2.4\times10^{-12}dcyc^{-1}$, which can not be explained in the angular momentum conservation case. The X-ray observation of other authors implies the accretion rate as about $3.6\times10^{-9} M_{\odot}yr^{-1}$ during high state. Those indicate the minimal distance of the decoupling point to the COM is about $0.81a$ which implies the mass lost from the secondary. And the magnetosphere covers about 42\% surface area of the secondary, which can cause the mass loss fraction of 58\% if the secondary goes through an intense stellar wind globally. This kind of mass loss fraction is very close to that deduced from losing mass through L2. The mass loss fraction is about  60 percent at least, so any model not taking into the effect of mass lost should be unrealistic.\\
2. The primary timescale of the flickering is about 2 minutes, which implies the threading point is about $30R_1$ above the surface of the WD. Within the high state, the flickering sticks out, and springs from the modulation of the accretion rate due to the wobbling of the threading point; also the data shows the intensity of flickering reduces when the system has an increasing accretion rate.\\
3. In the frame of two-pole model, the dominate accretion region is nearly face on to us near phase 0.7. Considering the continuing mass transfer during low state, this model predicts that the low-state LC should show intensity hump near phase 0.7, which contradicts the observation. The high-state LCs follow the shifting low-state LCs from phase 0 to phase 0.4, such large fraction of phase indicates the hump near phase 0.2 is caused by the WD itself rather than a tiny accretion region, which also does not conform to two-pole model. These two facts make the two-pole model less convincing.\\

\normalem
\begin{acknowledgements}
This work is supported by the Chinese Natural Science Foundation (Grant Nos. 11325315, 11611530685, 11573063, and 11133007), the Strategic Priority Research Program “The Emergence of Cosmological Structure” of the Chinese Academy of Sciences (Grant No.XDB09010202), and the Ministry of Education, Youth and Sports of the Czech Republic project LG15010. All observations in this paper were made at the 2.15-m Jorge Sahade telescope (JST) in Complejo Astron\'omico E1 Leoncito (CASLEO), San Juan, Argentina and the Danish 1.54-m Telescope at ESO La Silla, Chile. Finally, We are very grateful to the anonymous referee for an insightful report that has significantly improved the paper.
\end{acknowledgements} 

\bibliographystyle{raa}
\bibliography{MNHya_Ref}

\end{document}